\documentclass[conference]{IEEEtran}
\usepackage{makeidx,epsfig}
\usepackage{graphicx}
\usepackage{setspace,graphicx}
\usepackage{booktabs}
\usepackage{multirow}

\usepackage{url}

\begin{document}
\title{Nanotechnology: The New Features}
\author{\IEEEauthorblockN{Gang Wang}
\IEEEauthorblockA{Dept. of Computer Science and Engineering\\ University of Connecticut\\
Email: g.wang.china86@gmail.com $or$ gang.wang@uconn.edu}
}
\maketitle

\begin{abstract}
Nanotechnologies are attracting increasing investments from both governments and industries around the world, which offers great opportunities to explore the new emerging nanodevices, such as the Carbon Nanotube and Nanosensors. This technique exploits the specific properties which arise from structure at a scale characterized by the interplay of classical physics and quantum mechanics. It is difficult to predict these properties a priori according to traditional technologies. Nanotechnologies will be one of the next promising trends after MOS technologies. However, there has been much hype around nanotechnology, both by those who want to promote it and those who have fears about its potentials. This paper gives a deep survey regarding different aspects of the new nanotechnologies, such as materials, physics, and semiconductors respectively, followed by an introduction of several state-of-the-art nanodevices and then new nanotechnology features. Since little research has been carried out on the toxicity of manufactured nanoparticles and nanotubes,  this paper also discusses several problems in the nanotechnology area and gives constructive suggestions and predictions.
\end{abstract}
\IEEEpeerreviewmaketitle

\section{Introduction}
\label{Sec:Intr}
Nanotechnology, introduced almost half a century ago, is an active research area with both novel science and useful applications that has gradually established itself in the past two decades. Nanotechnology -- a term encompassing the science, engineering, and applications of submicron materials -- involves the harnessing of the unique physical, chemical, and biological properties of nanoscale substances in fundamentally new and useful ways. The economic and societal promise of nanotechnology has led to investments by governments and companies around the world. The complexities and intricacies of nanotechnology, still in the early stage of development, and the broad scope of these potential applications, have become increasingly important~\cite{RT_2014}. 

A nanometer is one-billionth of a meter. For example, a sheet of paper is about 100,000 nanometers thick; a single gold atom is about a third of a nanometer in diameter. Dimensions between approximately 1 and 100 nanometers are known as the nanoscale.

Nanotechnology is the understanding and control of matter at the dimensions between approximately 1 and 100 nanometers, where unique phenomena enable novel applications. Encompassing nanoscale science, engineering and technology, nanotechnology involves imaging, measuring, modeling, and manipulating matter at this length scale. The transformative and general purpose prospects associated with nanotechnology have stimulated more than 60 countries to invest in national nanotechnology research and development programs~\cite{RT_20111}. Unusual physical, chemical, and biological properties can emerge in materials at the nanoscale. These properties may appear dramatically different in important ways from the properties of bulk materials and single atoms or molecules~\cite{RT_2013}. Using structures designed at this extremely small scale, there exist opportunities to build materials, devices, and systems with nano-properties that can not only enhance existing technologies but also offer novel features with potentially far-reaching technical, economic, and societal implications~\cite{RT_2011}.

Nanotechnology products can be used for the design and processes in various areas. It has been demonstrated that nanotechnology has many unique characteristics, and can significantly fix the current problems which the non-nanotechnology faced, and may change the requirement and organization of design processes with its unique features~\cite{RT_20081}.

Nanotechnology deals with the production and applications at scales ranging from a few nanometers to submicron dimensions, as well as the integration of the resulting nanostructures into larger systems~\cite{HB_2010}. It also involves the investigation of individual atoms. Particularly, the conventional analytic aspects of nanotechnology must yield a certain synthetic approach, which is similar with non-nanotechnology. This action will be conducive to the creation of new functions exhibited by nanoscale structural units through their mutual interactions, even though these functionalities are not properties of the isolated units. We can use the term \textit{nanoarchitectonics} to express this innovation of nanotechnology~\cite{Narch_2012}: it is a technology system aimed at arranging nanoscale structural units, a group of atoms, molecules, or nanoscale functional components, into a configuration that creates a novel functionality through mutual interactions among those units.

These very small structures in nanoscale are intensely interesting for many reasons~\cite{White_2005}:
\begin{enumerate}
\item Many properties mystify us. For example, how do electrons move through organometallic nanowires?
\item They are challenging to make. For example, synthesizing or fabricating ordered arrays and patterns of nanoscale units poses fascinating problems.
\item Studying these structures leads to new phenomena, since many nanoscale structures have been inaccessible and/or off the beaten scientific track.
\item Nanostructures are in a range of sizes in which quantum phenomena, especially quantum entanglement and other reflections of the wave character of matter, would be expected to be important. 
\item The nanometer-sized, functional structures that carry out many of the most sophisticated tasks open up an exciting frontier of biology.
\end{enumerate}

Many nanotechnology advocates -- including business executives, scientists, engineers, medical professionals, and venture capitalists -- assert that in the longer term, nanotechnology, especially in combination with information technology, biotechnology, and the cognitive sciences, may deliver revolutionary advances~\cite{RT_2014}.

One set of considerations revolves around how nanotechnology is characterized, and how nanotechnology is understood as the emphasis moves toward the nano-era. In the discussion of the nano-era, there are divergent approaches to define and characterize the corresponding new features.

The rest of this paper is organized as follows: Section~\ref{Sec:Basic} briefly describes the basics of nanotechnology. Section~\ref{Sec:NF} presents the new features of nanotechnology. Section~\ref{Sec:ND} describe several nanotechnology devices. Section~\ref{Sec:Disc} discusses the problems and prediction of nanotechnology. Section~\ref{Sec:Conc} concludes this paper.

\section{Nanotechnology Basics}
\label{Sec:Basic}
Nanotechnology is the creation of materials and devices by controlling matter at the levels of atoms, molecules, and super-molecular structures~\cite{Roco_1999}, which means that it is the use of very small particles of materials to create new large-scale materials~\cite{Mann_2006}.

Nanotechnology whose form and importance are yet undefined is ``revolutionary nano": that is, technologies emerging from a new nanostructured material, or from the electronic properties of quantum dots, or from fundamentally new types of architectures -- based on nanodevices -- for use in computation and information storage and transmission. Nanosystems that use or mimic biology are also intensely interesting.

Even more thorough definitions and concepts of nanotechnology have been used by researchers in different areas as well, however, the key issue is the size of particles because the properties of materials are dramatically affected by the scale of the nanometer(\textit{nm}), $10^{-9}$ meter(\textit{m}). Actually, nanotechnology is not a new science or technology with current development as we spoke of above. The research on nanotechnology has been very active in the recent two decades for two reasons. One is the interesting features at the nanoscale, as we discussed in section~\ref{Sec:Intr}, and the other is that the development and application of nanotechnology rely on the rapid development of other related sciences and technologies, such as physics and chemistry.

According to ~\cite{HNC_05}, the subject of nanotechnology includes ``almost any materials or devices which are structured on the nanometer scale in order to perform functions or obtain characteristics which could not otherwise be achieved." 

To better understand the differences among various scales with regards to nanotechnology, Table~\ref{Tab:areas} shows the categories of the scales and their corresponding related areas~\cite{Bal_2005}.

\begin{table}[h]  \fontsize{8}{11}\selectfont
\caption{Particle Scales vs. Research Areas}
\label{Tab:areas}
\centering
\begin{tabular}{|p{2cm}||p{4cm}|}
\hline
 \textbf{\textit{Scales(meter)} }     &   \textbf{\textit{Research Areas(Not Inclusive)}} \\
\hline

\multirow{1}{*} {$10^{-12}$} & Quantum Mechanics \\
\hline

\multirow{4}{*}{$10^{-9}$} & \textbf{Nanomechanics}  \\
 & Molecular Dynamics \\
 & Molecular Biology \\
 & Biophysics \\
\hline

\multirow{3}{*}{$10^{-6}$} & Plasticity  \\
 & Elastictiy \\
 & Dislocation \\
\hline

\multirow{1}{*} {$10^{-3}$} & Mechanics of Materials\\
\hline

\multirow{1}{*} {$10^{-0}$} & Structural Analysis \\
\hline

\end{tabular}
\end{table}

Just because materials can be made into very small particles does not immediately mean that they have any practical use. However, the fact that these materials can be made at this nanoscale gives them the potential to have some interesting properties. Table~\ref{Tab:char} gives the characteristic lengths in solid-state science mode with respect to nanoscales~\cite{Char_2002}.

\begin{table}[h]  \fontsize{8}{11}\selectfont
\caption{Characteristic lengths in solid-state science model}
\label{Tab:char}
\centering
\begin{tabular}{|p{1.9cm}||p{3.5cm}|p{1.9cm}|}
\hline
 \textbf{\textit{Field} }     &   \textbf{\textit{Property}} &   \textbf{\textit{Scale Length}}\\
\hline

\multirow{3}{*}{Electronics} & Electronic Wavelength & 10 $\sim$ 100 nm  \\
 & Inelastic Mean Free Path & 1 $\sim$ 100 nm \\
 & Tunneling & 1 $\sim$ 10 nm \\
\hline

\multirow{3}{*}{Optics} & Quantum Well & 1 $\sim$ 100 nm  \\
 & Evanescent Wave Decay Length & 10 $\sim$ 100 nm \\
 & Metallic Skin Depth & 10 $\sim$ 100 nm \\
\hline

\multirow{2}{*}{Magnetics} & Domain Wall & 10 $\sim$ 100 nm  \\
 & Spin-flip Scattering Length & 1 $\sim$ 100 nm \\
\hline

\multirow{3}{*}{Superconductivity} & Cooper Pair Coherence Length & 0.1 $\sim$ 100 nm  \\
 & Meisner Penetration Depth & 1 $\sim$ 100 nm \\
\hline

\multirow{5}{*}{Mechanics} & Dislocation Interaction & 1 $\sim$ 1000 nm  \\
 & Grain Boundaries & 1 $\sim$ 10 nm \\
 & Crack Tip Radii & 1 $\sim$ 100 nm \\
 & Nucleation/Growth Defect & 0.1 $\sim$ 10 nm \\
 & Surface Corrugation & 1 $\sim$ 10 nm \\
\hline

\multirow{3}{*}{Supramolecules} & Kuhn Length & 1 $\sim$ 100 nm  \\
 & Tertiary Structure & 10 $\sim$ 1000 nm \\
 & Secondary Structure & 1 $\sim$ 10 nm \\
\hline

\multirow{1}{*}{Catalysis} & Surface Topology & 1 $\sim$ 10 nm  \\
\hline

\multirow{1}{*}{Immunology} & Molecular Recognition & 1 $\sim$ 10 nm  \\
\hline
\end{tabular}
\end{table}

According to quantum theory, materials at the nanoscale, between 1 nm and 250 nm, lie between the quantum effects of atoms, molecules and the bulk properties of materials. This nanoscale is called `No-Man's-Land' where many physical and electrical properties of materials are controlled by phenomena that have their own critical dimensions at the nanoscale.

Some `Nano' definitions used in this paper are listed below.

\begin{enumerate}
\item Cluster: A collection of units (atoms or reactive molecules) of up to about 50 units.
\item Colloids: A stable liquid phase containing particles in the 1-1000 nm range. A colloid particle is one such 1-1000 nm particle.
\item Nanoparticle: A solid particle in the range of 1-100 nm that could be noncrystalline, an aggregate of crystallites or a single crystallite.
\item Nanocrystal: A solid particle that is a single crystal in the nanometer range.
\end{enumerate}

With nanotechnology, scientists and engineers can influence, by being able to fabricate and control the structure of nanoparticles, the resulting properties and, ultimately, design materials to give designed properties. The electronic properties that can be controlled at this nanoscale are of great interest~\cite{May_2001}. The range of applications where the physical size of the particle can provide enhanced properties that are of benefit is extremely wide.

The science related to nanotechnology is new compared with other sciences. However, nanosized devices and objects have existed on earth as long as life. The exceptional mechanical performance of biomaterials, such as bones or mollusk shells, is due to the presence of nanocrystals of calcium compounds~\cite{HNC_06}. The history of technology suggests, however, that where there is smoke, there will eventually be fire; that is, where there is enough new science, important new technologies will eventually emerge~\cite{White_2005}.

Nanotechnology has changed and will continue to change our vision, expectations and abilities to control the materials and design world. These developments will definitely affect the semiconductor world and semiconductor materials. Recent major achievements include the ability to observe structure at its atomic level and measure the strength and hardness of microscopic and nanoscopic phases of composite materials.

The new features of nanotechnology materials and elements accordingly change nanotechnology usage, material force and resistance, as well as their related fields and designs. Therefore, it is essential and necessary to carefully study the new features of current nanotechnology.

\section{New Features}
\label{Sec:NF}
In this part, we will discuss the new features of nanotechnology from different scientific areas, such as materials, physics and information technologies(ITs). Although some new features in different areas may overlap in certain points, these features will display different properties or characteristics for specific areas. 

\subsection{Materials}
Much of nanoscience and nanotechnology is concerned with producing new or enhanced materials. Also, some nanotechnology-enabled products are already on the market and enjoying commercial success. These materials can behave quite differently at the nanoscale to the way they do in bulk. This is both because the small size of the particles dramatically increase surface area and therefore reactivity, and also because quantum effects start to become significant.

\subsubsection{3D Structure} 
Materials can be categorized by the overall dimensionalities of the structure and the class of compound. Many materials with nm dimensions in 1D have been commercially successful~\cite{MIN_00}.

Some recent novel developments include producing three-dimensional(3D)(particles), two-dimensional(2D)(monolayer films), one-dimensional(1D)(wires and tubes) and zero-dimensional(0D)(dots) for functional applications. This section will be concentrated on the developments and structures of 3D carbon particles.

Carbon nanostructures have been the focus of much interest and research since they were first observed in the mid-1980s~\cite{DON_03}. The football-shaped Buckminsterfullerene($C_{60}$) and its analogs show great promise as lubricants and, thanks to their cage structures, as drug delivery systems, as well as in electronics. The same graphite sheet structure, which allows electrical conductivity, was discovered in the early 1990s~\cite{DON_04}.

\begin{figure}
  \centering
  \includegraphics[width=5cm]{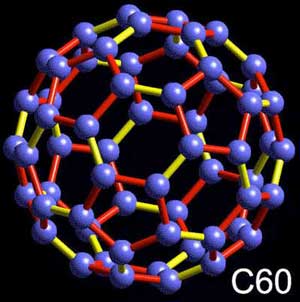}
  \caption{Carbon C60 A Beautiful Molecule~\cite{Web_C02}}
  \label{Fig:C60}
\end{figure}

Fullerenes consist of 20 hexagonal and 12 pentagonal rings as the basis of an icosahedral symmetry closed cage structure. Each carbon atom is bonded to three others. The $C_{60}$ molecule has two bond lengths - the 6:6 ring bonds can be considered as ``double bonds" and are shorter than the 6:5 bonds. $C_{60}$ is not "super aromatic" as it tends to avoid double bonds in the pentagonal rings, resulting in poor electron delocalization. As a result, $C_{60}$ behaves as an electron deficient alkene, and reacts readily with electron rich species. The geodesic and electronic bonding factors in the structure account for the stability of the molecule. In theory, an infinite number of fullerenes can exist, their structure based on pentagonal and hexagonal rings, constructed according to rules for making icosahedra~\cite{Web_C01}. Fig.~\ref{Fig:C60} shows the 3D structure of the fullerenes~\cite{Web_C02}.

\subsubsection{Surface Ratio}
In many sub-fields of nanotechnology, advances in structured materials occur both by evolutionary development of technologies and by revolutionary discoveries that generated new approaches to materials synthesis. As the particle size approaches to the 10 $\sim$ 100 nm range, the surface to volume ratio increases and properties become size dependent.

When the dimensions of materials are decreased from macrosize to nanosize, significant changes in electronic conductivity, optical absorption, chemical reactivity, and mechanical properties occur. With the decrease in size, more atoms are located on the surface of the particle. Also, these particles can be considered as nanocrystals and the atoms within the particle are perfectly ordered or crystalline.

\begin{figure}
  \centering
  \includegraphics[width=9cm]{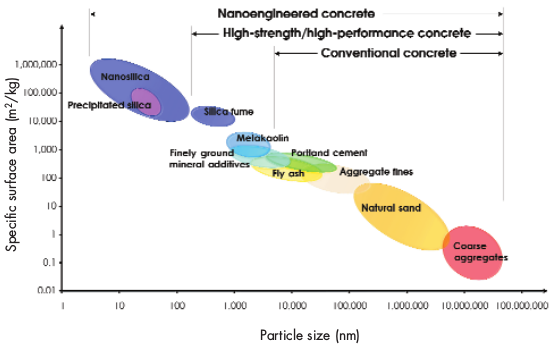}
  \caption{Particle-size and specific-surface-area scale related to concrete materials~\cite{Web_03}.}
  \label{Fig:surf}
\end{figure}

\begin{figure}
  \centering
  \includegraphics[width=6cm]{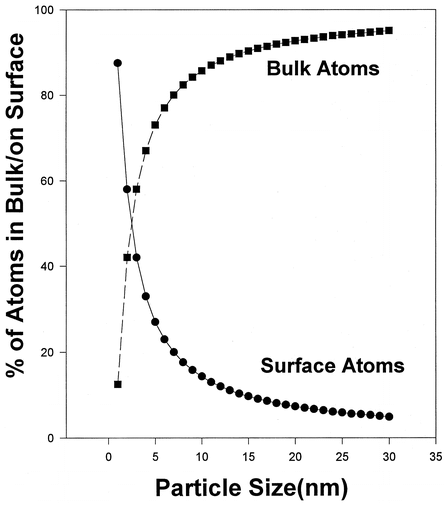}
  \caption{Calculated surface to bulk ratios for solid metal particles vs size~\cite{NAS}.}
  \label{Fig:surf1}
\end{figure}

Nanoparticles have a remarkable surface area, as shown in Fig.~\ref{Fig:surf}. The calculated surface to nanoparticles bulk ratios for solid metal particles \textit{vs.} size is shown in Fig.~\ref{Fig:surf1}. The surface area imparts a serious change of surface energy and surface morphology. All these factors alter the basic properties and the chemical reactivity of the nanomaterials~\cite{HB_2010}~\cite{HNC_07}~\cite{HNC_10}. The change in properties causes improved catalytic ability, tunable wavelength-sensing ability and better-designed pigments and paints with self-clean and self-healing features~\cite{HNC_00}.

The Laplacian (a differential operator given by the divergence of the gradient of a function on Euclidean space) pressure, due to surface energy and the atomic structure of the surfaces, impacts density, phase transition temperatures, interface potential, and those properties that depend upon them. However, when the particle size is below 10 nm, the quantum effects dominate. 

A number of research groups, notably UC Berkeley and MIT, developed synthetic strategies to produce particles of semiconductor and metal nanocrystals with particle diameters in the range of 1 $\sim$ 50 nm. The new methods involve the injection of molecular precursors into hot organic surfactants and yield narrow size distributions, good size control and good crystallinity of dispersable nanocrystals~\cite{MIN_05}~\cite{MIN_06}~\cite{MIN_07}. In this size range, the optical absorption of compounds is a sensitive function of particle size.

Exciting extensions of surfactant-mediated growth take advantages of the fact that absorption of surfactants is dependent on the atomic structure of the surface. The shapes of nanodots are determined by differences in the surface energies of the terminating atomic planes. By designing surfactants that preferentially absorb on specific crystal planes and by using more than one surfactant simultaneously, the direction dependence of growth rate can be tailored. One can imagine these as the basis of 3D functional structure as we mentioned above.

\subsubsection{Quantum Effects}
Quantum mechanics is a fundamental branch of physics which deals with physical phenomena at nanoscopic scales, where the action is on the order of the Planck constant. The name derives from the observation that some physical quantities can change only in discrete amounts (Latin quanta), and not in a continuous (cf. analog) way~\cite{Web_04}.

Several phenomena become pronounced as the size of the system decreases. These include statistical mechanical effects, as well as quantum mechanical effects, for example, the ``quantum size effect" where the electronic properties of solids are altered with great reductions in particle size. This effect does not come into play by going from macro to micro dimensions. However, quantum effects can become significant when the nanometer size range is reached, typically at distances of 100 nanometers or less, the so-called quantum realm. Additionally, a number of physical (mechanical, electrical, optical, etc.) properties change when compared to macroscopic systems. One example is the increase in surface area to volume ratio altering mechanical, thermal and catalytic properties of materials. Diffusion and reactions at the nanoscale, nanostructures materials and nanodevices with fast ion transport are generally referred to as nanoionics. Mechanical properties of nanosystems are of interest in nanomechanics research. The catalytic activity of nanomaterials also opens potential risks in their interaction with biomaterials~\cite{Web_05}.

At the nanoscale, quantum confinement effects dominate the electrical and optical properties of systems~\cite{QPI}. Much interest is also focused on quantum dots, which are semiconductor nanoparticles that can be 'tuned' to emit or absorb particular colors of light for use in solar energy or fluorescent biological labels.

Electrons localized in a quantum dot by a confinement potential occupy atomic like states with discrete energy levels. Therefore, a quantum dot with confined electrons is called the artificial atom~\cite{MOE_01}. In the electrostatic or gated quantum dots~\cite{MOE_02}~\cite{MOE_05}, the confinement potential results from the external voltages, applied to the electrodes, and band offsets. The confinement potential is vary sensitive to the voltages applied as well as the parameters of the nanostructure, in particular, the geometry of the nanodevice and doping. The electronic properties of the nanodevice are also determined by the confinement potential. Therefore, the knowledge of the realistic profile of this potential is important for a design of the nanodevice with the required electronic properties and for a theoretical description of the confined electron states~\cite{MOE}.

Also, quantum dots are being developed as labels in medical imaging and have potential in nano-opto electronics.

\subsection{Physicals}
Nanoparticles often have their own physical and chemical properties that are very different from the same materials at larger scales.
The properties of nanoparticles depend on their shape, size, surface characteristics and inner structure. They can change in the presence of certain chemicals.
The composition of nanoparticles and the chemical processes taking place on their surface can be very complex.
Nanoparticles can remain free or group together, depending on the attractive or repulsive interaction forces between them~\cite{Web_06}.

\subsubsection{Self-Assembly}
\begin{figure}
  \centering
  \includegraphics[width=7cm]{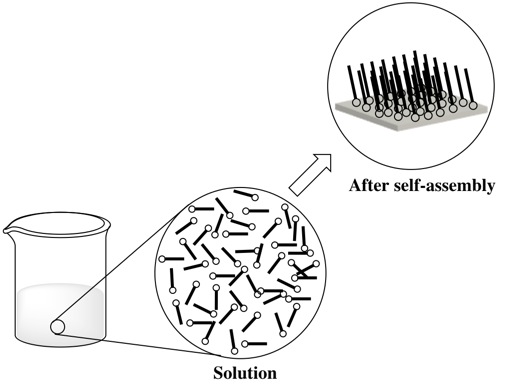}
  \caption{Self-assembly processes of Nanoparticles~\cite{Web_08}.}
  \label{Fig:Self}
\end{figure}

Self-assembly is a phenomenon where the components of a system assemble themselves spontaneously via an interaction to form a larger functional unit. This kind of spontaneous organization can be due to direct specific interaction and/or indirectly through their environment. Due to increasing technological advancements, the study of materials on the nanometer scale is becoming more important. The ability to assemble nanoparticles into a well-defined configuration in space is crucial to the development of electronic devices that are small but can contain plenty of information. The spatial arrangements of these self-assembled nanoparticles can be potentially used to build increasingly complex structures leading to a wide variety of materials that can be used for different purposes~\cite{Web_07}~\cite{DSA}~\cite{SAO}. Fig.~\ref{Fig:Self} shows the self-assembly processes of nanoparticles.

With the continuous development of nanotechnology, the possibility for the bottom-up production of nanoscale materials may result in some kind of self-assembly of structures similar to the self-assembly of phospholipid bilayers that resembles cellular membranes. On the basis of current knowledge, however, the spontaneous formation of artificial living systems through self-assembly and related processes, suggested by some prominent commentators, is considered highly improbable. The combination of self-replication with self-perpetuation in an engineered nanosystem is extremely difficult to realize on the basis of current scientific knowledge.

By the controlled self-assembly and self-organization of molecular compounds and supramolecular entities, respectively, it should be possible to design rationally and to construct precisely nanoscale molecular devices with switching properties~\cite{TAA}.

Nanotechnology is dependent on nanostructures that require creation and characterization. Two fundamentally different approaches for the controlled generation of nanostructures have evolved. On one hand, there is growth and self-assembly, from the bottom up, involving single atoms and molecules. On the other hand, there is the top-down approach in which the powerful techniques of lithography and etching start with large uniform pieces of material and generate the required nanostructures from them. Both methods have inherent advantages. Top down assembly methods are currently superior for the possibility of interconnection and integration, as in electronic circuitry. 
Bottom-up assembly is very powerful in creating identical structures with atomic precision, such as the supramolecular functional entities in living organisms. In many different fields of nanoscale science, e.g. the production of semiconductor quantum dots for lasers, the production of nanoparticles by a self organization, and the generation of vesicles from lipids, self-organization is used for the generation of functional nanometre-sized objects. To date, man made self-organized structures ~\cite{NPA} remain much simpler than nature’s complex self-organized processes and structures. 

As noted above, there is also no reason to believe that processes of self-assembly, which are scientifically very important for the generation of nanoscale structures, could lead to uncontrolled self perpetuation~\cite{Web_06}. 

\subsubsection{Magnetic} 
Magnetic nanoparticles are a class of nanoparticles which can be manipulated by using magnetic field gradients. Such particles commonly consist of magnetic elements such as iron, nickel and cobalt and their chemical compounds. While nanoparticles are smaller than 1 micrometer in diameter (typically 5–500 nanometers), the larger microbeads are 0.5–500 micrometer in diameter. Magnetic nanoparticle clusters which are composed of a number of individual magnetic nanoparticles are known as magnetic nanobeads with a diameter of 50–200 nanometers~\cite{Web_09}~\cite{MPO1}.
The physical and chemical properties of magnetic nanoparticles largely depend both on the synthesis method and chemical structure. In most cases, the particles range from 1 to 100 nm in size and may display superparamagnetism~\cite{MNS}. We will talk about the quantum tunneling in the magnetic nanoparticles below.

One of the fascinating properties of magnetic nanoparticles is the reduction from multidomains to a single domain as the particle size reduces to some limit values. Besides the vanishing of magnet hysteresis and the large reduction of the coercive field for nanoparticles, the macroscopic quantum tunneling of the magnetic moment becomes its non-analyticity in the ground state energy of the infinite lattice system~\cite{REN_211}. Unusual electronic and magnetic characteristics are prevalent at nanozero temperatures such as the metal-insulator transition in transition metal oxides~\cite{REN_212}, non-Fermi-liquid behavior of highly correlated \textit{f}-electron compounds~\cite{REN_213}~\cite{REN_214}, abnormal symmetry states of high-$T_c$ superconducting heterostructures. The investigation of the remarkable properties of these systems has attracted great attention by researchers in condensed matter physics. The physics underlying the quantum phase transitions described above is quite involved and in many cases, has not been completely understood so far.

A surface spin-glass layer is proven to be ubiquitous in magnetic nanoparticles at low temperatures~\cite{REN_218}. A larger surface to volume ratio of the small nanoparticles implies a stronger surface anisotropic field to frustrate and disorder the inner spins, causing quantum tunneling at higher temperatures~\cite{REN_219}~\cite{REN_220}~\cite{REN_221}.

At a low temperature, magnetic viscosity of these systems shows a constant value below a finite temperature reflecting the independence of thermally over-barrier transitions and is the signature of quantum tunneling of magnetization. However, at high temperatures, single-domain magnetic nanoparticles are thermally free to orient their spin directions and exhibit superparamagnetic properties. The superparamagnetic state is blocked as the temperature lowers down to enhance the exchange interactions between particles.

\subsubsection{Dielectric}
A dielectric material, or dielectric, is an electrical insulator that can be polarized by an applied electric field. When a dielectric is placed in an electric field, electric charges do not flow through the material as they do in a conductor, but only slightly shift from their average equilibrium positions causing dielectric polarization. As materials considered for inclusion in nanodevices are designed for more complex behavior, dielectric properties have become increasingly important. Characterization of nonlinear properties, such as piezoelectric, ferroelectric, and ferromagnetic responses is now critical.

The dielectric constant, the response function of the measured external field, the displacement, to the local electric filed, closely relates to the conductivity and optical properties of materials. The dielectric constants of metallic nanoparticles in microwave frequency range have rarely been reported~\cite{PPO_201}. The high microwave field absorption of the metallic particles involves using the conventional method of inserting a powder-pressed thin disk in a microwave guide to determine the dielectric constant by measuring the attenuation and phase delay of the penetrating wave, which cannot be used~\cite{PPO}.

The electrical and magnetic properties of numerous nanomaterials are completely different from those of their bulk counterparts. Changes in dielectric properties were attributed to changes in particle size, shape, and boundaries~\cite{MNO_23}~\cite{MNO_24}. The modified dielectric properties were used as capacitors, electronic memories, and optical filters. Materials exhibiting a giant dielectric constant have already been reported elsewhere~\cite{MNO_26}~\cite{MNO_27}. The high dielectric permittivity and the low loss factors over a wide frequency range are always of a great interest ~\cite{MNO}~\cite{MNO_28}.

Dielectric constants specify the response to the dipole displacement in an externally applied field in terms of ion and electron motion. Incident electromagnetic (EM) fields of different frequencies cause different responses from ions and electrons. As the size of the metal films or particles declines, the mean free path becomes constrained by surface scattering. The conductivity of metallic nanoparticles decreases as the particle size decreases and behaves as non-conducting below a critical size and temperatures.

The magnitude of the real part of dielectric constants for metallic nanoparticles decreases with a decreasing particle size~\cite{PPO}, suggesting that the particles become less conducting as the particle size decreases. The microwave absorption depends on the shape and size distribution, making it extremely difficult to determine the imaginary part. The darkish appearance of many different metallic nanoparticles illustrates that the measured dielectric constants, even of silver and iron nanoparticles, are close in proximity.


\subsection{Semiconductors}
\label{Sec:NF_Semic}
A semiconductor material has an electrical conductivity value falling between that of a conductor, such as copper, and an insulator, such as glass. Semiconductors are the foundation of modern electronics. Semiconducting materials exist in two types - elemental materials and compound materials~\cite{SPA}.

\begin{figure}
  \centering
  \includegraphics[width=7cm]{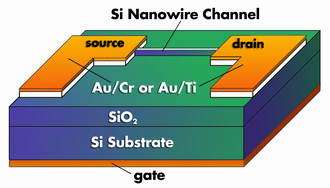}
  \caption{Nanoscaled transistor can be fabricated by inserting nanowires (or single molecule) between source and drain electrodes mounted on a silicon/silicon dioxide support~\cite{CNM}.}
  \label{Fig:trans}
\end{figure}

The transition from devices relying on collections of molecules to unimolecular devices requires the identification of practical methods to contact single molecules. This fascinating objective demands the rather challenging miniaturization of contacting electrodes to the nanoscale. A promising approach to unimolecular devices relies on the fabrication of nanometer-sized gaps in metallic features followed by the insertion of individual molecules or nanomaterials between the terminals of the gap. The strategy permits the assembly of nanoscaled three-terminal devices equivalent to conventional transistors. Fig.~\ref{Fig:trans} shows a nanoscaled transistor which can be fabricated by inserting nanowires.

The selected examples of tubes, wires or quantum dots in this section only hint at the range of materials from which nanostructures are made. The summary in Table~\ref{Tab:Nano_materils}  provides some perspective of the diversity that is possible today. 

\begin{table}[h]  \fontsize{8}{11}\selectfont
\caption{Nanostructured materials}
\label{Tab:Nano_materils}
\centering
\begin{tabular}{|p{2cm}|p{4cm}|}
\hline
 Nano tubes     &  Carbon, VxOy, SnO$_{2}$ , InAs, GaAs, GaN, Co$_{3}$O$_{4}$ , BN, WS$_{2}$ , ZrO$_{2}$ , MoS$_{2}$ , H$_{2}$Ti$_{3}$O$_{7}$ , polypyrrole, peptides, metallo porphyrin, SiO$_{2}$ , Cu  \\
\hline

 Nano wires     &  Si, In, InP, InAs, MgO, MoSe, GaN, Ga$_{2}$O$_{3}$ , ZnO, SnO$_{2}$ , TiO$_{2}$ ,
Pt, Au, Ag, Ni, Cu, Bi, Co, Pb, LiMnO$_{2}$ , CdTe, LiNiO$_{2}$ , CdS, B, PbSe, FeCo, FeNi, CoPt, BN, ZnS, ZnSe, CdSe, SiGe, ErSi$_{2}$, DySi$_{2}$, polyanaline  \\
\hline

 Nano dots     &   GaAs, InP, Si, InAs, CdS, CdSe, TiO$_{2}$, ZnS, Fe$_{2}$O$_{3}$, MnO$_{4}$, Cr$_{2}$O \\
\hline

\end{tabular}
\end{table}

\subsubsection{Nanotubes}
\begin{figure}
  \centering
  \includegraphics[width=7cm]{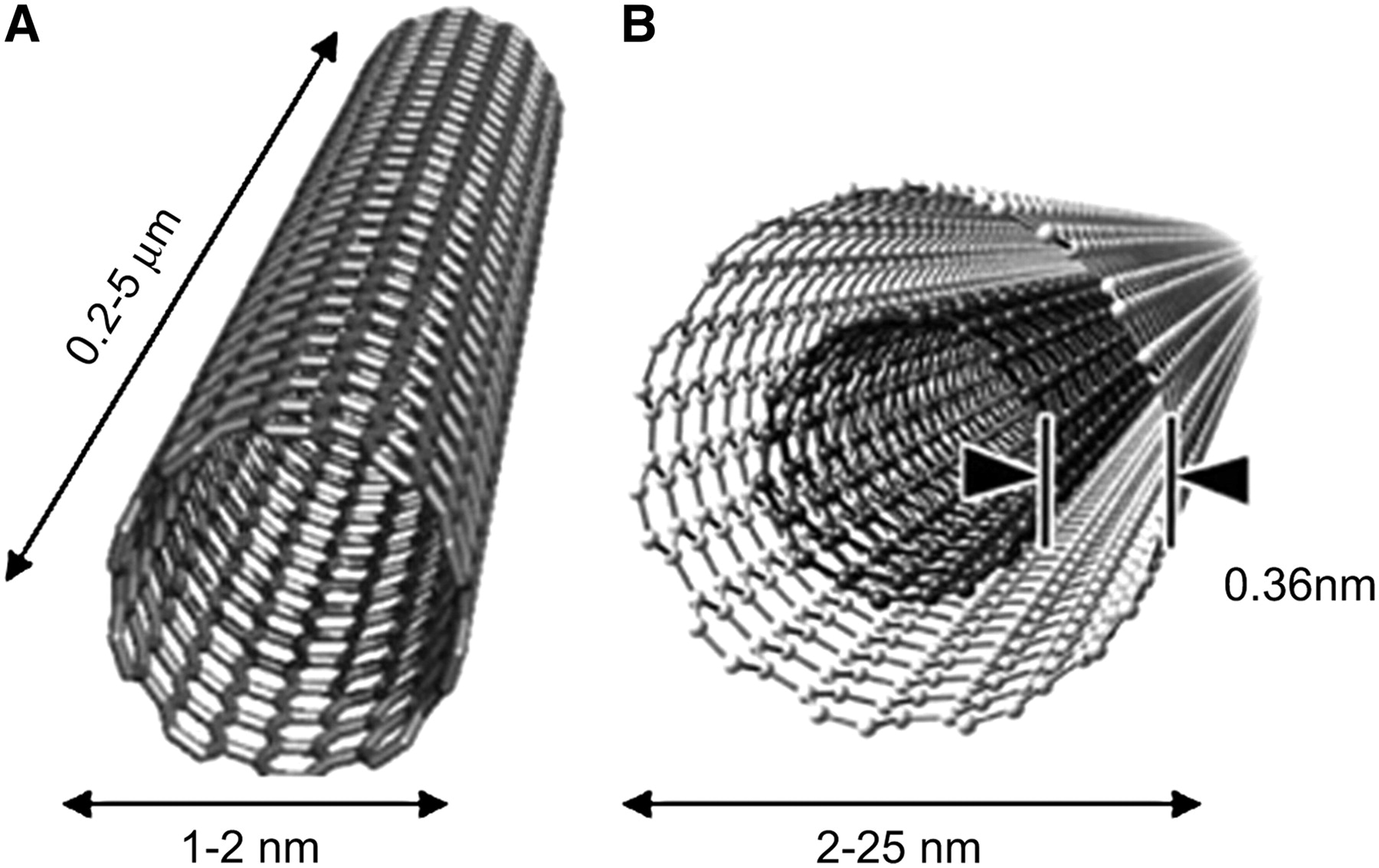}
  \caption{Conceptual diagram of single-walled carbon nanotube (SWCNT) (A) and multiwalled carbon nanotube (MWCNT) (B) delivery systems showing typical dimensions of length, width, and separation distance between graphene layers in MWCNTs~\cite{NTubes}.}
  \label{Fig:NTubes}
\end{figure}

A nanotube is a nanometer-scale tube-like structure. Semiconductor nanotubes are a natural candidate for three terminal nanodevices. The nanotube is positioned to bridge two metal electrodes, which as the source and drain of the field-effect transistors(FETs). The silicon wafer is used as a back gate. These devices behave as unipolar \textit{p}-type FETs with on/off current switching ratios of $\sim 10^{5}$. However, the first devices had a high contact resistance($>$1 M $\Omega$) which led to a low conductance $\sim 10^{-9}$ A/V. Subsequent improvements in the processing result in decreases in contact resistance by three orders of magnitude and increase conductance by two orders of magnitude. These nanotubes were \textit{p}-type. \textit{n}-type nanotubes can be made by doping\cite{MIN_27} or annealing in a vacuum~\cite{MIN_28}. Several groups have demonstrated complex devices using a combination of tubes or tube/metal interfaces. Fig.~\ref{Fig:NTubes} shows a conceptual diagram of single-walled carbon nanotube and multiwalled carbon nanotube.

It was soon learned that nanotubes have high strength and modulus, have interesting thermal conductivity, and the electrical properties are sensitive to surface adsorption. This combination of properties suggests applications that range from reinforced polymer composites, chemical sensors, field emission displays, drug delivery devices, thermal management systems, and SPM tips. Consequently, activity in this area has increased dramatically. The primary challenge to use of nanotubes is that current synthesis processes cannot produce tubes with predefined lengths or properties. Much effort is being expended on finding schemes for selection and/or separation.

Although the first nanotubes were carbon based, this by no means defines a fundamental limitation. The general appeal of nanotubes from more complex materials has motivated synthesis of a wide range of compounds in this geometry. This area is growing rapidly.

Carbon is a unique light atom that can form one-, two-, or threefold string chemical bonds. The planar threefold configuration forms graphene planes that can, under certain growth conditions, adopt a tubular geometry. Properties of carbon nanotubes may change dramatically depending on whether single-wall carbon nanotubes(SWNT) or Multiwall Nanotubes(MWNT) are considered. We will consider several properties of carbon nanotubes below~\cite{HB_2010}.

\begin{enumerate}
\item Variability of carbon nanotube properties.\\ 
Properties of MWNTs are generally not much different from that of regular polyaromatic solids, and variations are then mainly driven by the textural type of MWNTs considered and the quality of the nanotexture, both of which control the extent of anisotropy. However, the properties for SWNTs may change dramatically depending on whether single SWNT or SWNT ropes are involved. Note that: The following description will emphasize SWNT properties, as far as their original structure often leads to original properties with respect to that of regular polyaromatic solids.
\item General properties. \\
SWNT-type carbon nanotube diameters fall in the nanometer range and can be hundreds of micrometers long. SWNTs are narrower in diameter than the thinnest line able to be obtained by electron beam lithography. While the length of SWNTs can be macroscopic, the diameter has a molecular dimension. As a molecule, properties are closely influenced by the way atoms are displayed along the molecule direction. The physical and chemical behaviors of SWNTs are therefore related to their unique structural features.
\item SWNT adsorption properties. \\
The very high surface area, as we talked about before, yields many interesting features. Theoretical calculations have predicted that the molecule adsorption on the surface or inside of nanotube bundle is stronger than that on an individual tube. 
\begin{enumerate}
\item Accessible SWNT Surface~\cite{HB_3122}~\cite{HB_3123}~\cite{HB_3124}.
\item Adsorption Sites and Binding Energy of the Adsorbates~\cite{HB_3127}~\cite{HB_3133}
\end{enumerate}
\item Transport properties.\\
The narrow diameter of SWNTs has a strong influence on its electronic excitation due to its small size compared to the characteristic length scale of low energy electronic excitation. Combined with the particular shape of the electronic band structure of graphene, carbon nanotubes are ideal quantum wires.

\item Mechanical properties. \\
While tubular nano-morphology is also observed for many two-dimensional solids, carbon nanotubes are unique through the particularly strong three-folded bonding of curved graphene sheet, which is stronger than in diamond as revealed by their difference in C--C bond length. This makes carbon nanotubes --SWNTs or c-MWNTs -- particular stable against deformations.

\item Reactivity.\\
The chemical reactivity of graphite, fullerenes, and carbon nanotubes exhibits some common features. Like any small object, carbon nanotubes have a large surface with which they can interact with their environment. It is worth noting, however, that the chemistry of nanotubes differs from that of regular polyaromatic carbon materials because of their unique shape, small diameter, and structural properties.
\end{enumerate}

Those above properties make nanotubes much more suitable to nanosemiconductors circuits compared with CMOS circuits.

\subsubsection{Nanowires}
Nanotubes of the length longer than 1 $\mu m$ are usually called nanowires or nanofibers. Nanowires are especially attractive for nanoscience studies as well as for nanotechnology applications. They can be prepared by physics, chemistry or the mixture to produce metallic wires, and semiconductors~\cite{RNM_57}~\cite{RNM_60}. Nanowires, compared to other low dimensional systems, have two quantum confined directions while still leaving one unconfined direction for electrical conduction. This allows them to be used in applications which require electrical conduction, rather than tunneling transport. Because of their unique density of electronic states, nanowires in the limit of small diameters are expected to exhibit significantly different optical, electrical, and magnetic properties from their bulk 3D crystalline counterparts. 

Increased surface areas, very high density of electronic states and joint density of states near the energies of their van Hove singularities, enhanced excitation binding energy, diameter-dependent bandgap, and increased surface scattering for electrons and phonons are just some of the ways in which nanowires differ from their corresponding bulk materials. Yet the sizes of nanowires are typically large enough($>$1 nm in the quantum confined direction) to have local crystal structures closely related to their parent materials, thereby allowing theoretical predictions about their properties~\cite{HB_2010}.

Due to the enhanced surface-to-volume ratio in nanowires, their properties may depend sensitively on their surface condition and geometrical configuration. Even nanowires made of the same material may possess dissimilar properties due to differences in their crystal phase, crystalline size, surface conditions, and aspect ratios, which depend on the synthesis methods and conditions used in their preparation. Below are listed several novel properties of nanowires.

\begin{enumerate}
\item Transport Properties

The study of nanowire electrical transport properties is important for nanowire characterization, electronic device applications, and investigation of unusual transport phenomena arising from one-dimensional quantum effects. Important factors that determine the transport properties of nanowires include the wire diameter, material composition, surface conditions, crystal quality, and the crystallographic orientation of the wire axis, which is important for materials with anisotropic materials parameters, such as the effective mass tensor, the Fermi surface, or the carrier mobility. 

Electronic transport phenomena in low-dimensional systems can be roughly divided into two categories: ballistic and diffusive transport. Ballistic transport phenomena occur when electrons travel across the nanowire without any scattering; however, the transport is in the diffusive regime, and the conduction is dominated by carrier scattering within the wires due to phonons, boundary scattering, lattice and other structural defects, and impurity atoms.

a). Conduction Quantization in Metallic Nanowires:

The phenomenon of conductance quantization occurs when the diameter of the nanowire is comparable to the electron Fermi wavelength, which is on the order of 0.5 nm for most metals~\cite{HB_4125}. Most conductance quantization experiments up to the present have been performed by joining and separating two metal electrodes.

b). Diameter-dependent Properties of Semiconductor Nanowires:

The electronic transport behavior of nanowires may be categorized based on the relative magnitude of three length scales: carrier means free path, the de Broglie wavelength of electrons and wire diameter. For different relations among the three length scales, nanowires exhibit different transport properties and show some extend diameter-dependent properties~\cite{HB_2010}.  
Transport properties of nanowires in the classic finite size and quantum size regimes are highly diameter-dependent.

c). Thermoelectric Properties:

Nanowires are predicted to be promising for thermoelectric applications~\cite{HB_4134}~\cite{HB_4147}, due to their novel band structure compared to their bulk counterparts and the expected reduction in thermal conductivity associated with enhanced boundary scattering. Due to the sharp density of states at the 1D subband edges, nanowires are expected to exhibit enhanced Seebeck coefficients compared to their bulk counterparts.

d). Thermal Conductivity of Nanowires:

The thermal conductivity of small homogeneous nanowires may be more than one order of magnitude smaller than in the bulk, arising mainly from strong boundary scattering effects~\cite{HB_4154}. And, phonon confinement effects may eventually become important at still smaller diameter nanowires.

\item Optical Properties

Optical methods provide an easy and sensitive tool for measuring the electronic structure of nanowires since optical measurements require minimal sample preparation and the measurements are sensitive to quantum effects.

Phonons in nanowires are spatially confined by the nanowire cross-sectional area, crystalline boundaries and surface disorder. These finite size effects give rise to phonon confinement, causing an uncertainty in the phonon wave vector, which typically gives rise to a frequency shift and a lineshape broadening. Since zone center phonons tend to correspond to maxima in the phonon dispersion curves, the inclusion of contributions from a broader range of phonon wave vectors results in both a downshift in frequency and an asymmetric broadening of the Raman line that develops a low-frequency tail.

\end{enumerate}

\subsubsection{Quantum Dots}

The study of quantum dots (QD) has a longer history, arising as it does out of the semiconductor field of quantum wells, heterostructures, low dimensional electron gasses, etc~\cite{MIN_00}.
Semiconductor quantum dots with tunable optical emission frequencies due to the quantum size confinement present the utmost challenge and point of culmination of semiconductor physics. A modified Stranski-Krastanow growth method driven by self-organization phenomena at the surface of strongly strained heterostructure driven has been realized~\cite{RNA_24}.

Quantum dots can be many things theoretically, but the initial products that incorporate quantum dots are small grains (a few nanometers in size) of semiconductor materials~\cite{White_39}~\cite{White_40}. These grains are stabilized against hydrolysis and aggregation by coating with a layer of zinc oxide and a film of organic surfactant, technologies already familiar to the chemical industry in making paints and washing powders. These first semiconductor quantum dots are fluorescent -- they emit colored light when exposed to ultraviolet excitation -- and are being tested in displays for computers and mobile telephones, and as inks. These materials are interesting for several reasons: one is that they do not photobleach(that is losing their color on exposure to light); a second is that a single manufacturing process can make them in a range of sizes, and thus, in a single process, in all colors. Their applications in biology illustrate the difficulties in introducing a new technology~\cite{White_2005}.

All semiconductor devices incorporate a crucial ingredient for their proper functioning. The great interest in understanding the properties of these impurity containing systems comes from the fact that the impurity modifies the energy levels of the materials and in turn affects their electronic and optical properties~\cite{IBS_01}~\cite{IBS_02}. Thus, these systems have potential use in electro-optical devices~\cite{IBS_03}. A consequence of the strong confinement of the impurity states in quantum dots is that their electronic structures collapse to a series of discrete levels, contrary to the continuous source and drain associated with bulk semiconductors, or to their higher dimensional neighbors such as quantum wells and quantum wires~\cite{IBS_00}.

The study of bound impurity states in such structures is recently considered to be a subject of fundamental interest~\cite{IBS_04}~\cite{IBS_06}~\cite{IBS_07}~\cite{IBS_13}. There are ample investigations that highlight the influence of the mechanism and control of dopant incorporation, as well as impurity location in characterizing several properties of the quantum dots of nanodevices~\cite{IBS_14}~\cite{IBS_17}~\cite{IBS_23}. The effects of impurity need to be accessed on structure, electron density, and information entropy, etc., in case device level applications based on dot atoms are envisaged. 

It needs to be mentioned that quantum dots are now realizable in various shapes and sizes and device applications. As the physical dimensions of the dot approach the nanometer scale, size effects begin to play an important role, leading to drastic changes in measured properties~\cite{IBS_31}. As fabrication processes improve, control of dot size is enhanced. In the last few years, semiconductor quantum dots with tunable size have attracted a great deal of attention, particularly in the 1.3$\sim$1.55 $\mu m$ range of optical communications~\cite{IBS_33}~\cite{IBS_34}. The location of the impurity center together with confining fields could create a geometry where the dot size would display significant sensitivity in modulating the energy levels. Such an isolated impurity (electron-type) doped quantum dot structure could be a test case serving as representative of experimentally realizable ones. 

Researchers and scientists are interested in the nanoscale, because when many materials get down to these tiny sizes, they start to behave differently and novel properties emerge. Sometimes the material becomes explosive or its melting point changes or a new property is revealed. These novel properties are mostly due to changes in size and scale and the physics ‘rules’ that govern materials at the nanoscale.
Many novel properties are emerging as materials are being reduced from macroscale to nanoscale. This change in the properties of materials is leading to the creation of new and enhanced nanomaterials. Nanoscale materials like nanoparticles and nanofibres have an exciting future in a wide range of high-tech applications.

\section{NanoDevices}
\label{Sec:ND}
A major issue in nanoscale research is how the scientific paradigm changes will be translated and implemented into novel technological processes. Nanoparticle systems, including nano-clusters, nanotubes, nanostructured particles, and other three-dimensional nanostructures in the size range between 1 and 100 nm are usually seen as the tailored precursors for nanostructures materials and corresponding nanodevices.

Different application areas together promote the development of nanodevices, such as information and communication technologies, automotive, aerospace, energy, medical/pharmaceutical, chemicals and advanced materials,  textiles etc. In this paper, we mainly focus on the various nano-transistors and nanosensors used in computation and information storage and transmission areas.

\subsection{MOS Transistors}
The ever progressing and seemingly unstoppable miniaturization of MOS (Metal Oxide Semiconductor) transistors becomes the essential factor responsible for the continuous progress of nanotechnology. MOS transistors with channel lengths of around 100 nm have already been introduced in lots of semiconducting areas, such as the production of memory modules and microprocessors. Now newly developed silicon transistors with the channel length down to 18 nm have been popularly used in the fabrications of MOS technology, according to the ITRS surveys~\cite{ITRS_2013}.

However, MOS technology is not just miniaturizing the size of transistors and special processes  are needed to accomplish the transition from micrometer scale to the nanometer regime. Following are the essential process steps for transition~\cite{NAN_2005}
\begin{enumerate}
\item[$\bullet$] Adjustment of the gate oxide thickness to a few nanometers.
\item[$\bullet$] Reduction of the doping depths to a few nanometers.
\item[$\bullet$] Optimization of the spacer width and of the LDD (Lightly Doped Drain) doping.
\item[$\bullet$] Optimization of the channel doping.
\item[$\bullet$] Introduction of special implantation (such as pocket implantation).
\end{enumerate}

For transistors with the channel lengths below 100 nm, parasitic short channel effects become increasingly dominant and are difficult to reduce with the usual countermeasures. Therefore, measures for transitions, such as a further reduction of the gate oxide thickness or the decrease of all doping depths, are both technologically and physically limited for mass fabrications.

While the electrical characteristics of MOS transistors such as slop and switching speeds have been improved largely in the recent years by the progressive reduction of the transistor dimensions regarding current manufacturing technologies, a rather opposite trend is to be expected for the sub-100 nm transistors. However, dynamic investigations show a trend that the switching speed of sub-100 nm MOS transistors does not increase by the amount that is generally expected. The possible reasons are partly due to the increasing doping gradients which lead to increasing parasitic capacitance of transistors. Analyses by a large number of independent scientists show, however, that in the future the delay time in the signal lines of the microchip will be one of the dominate effects on the electrical characteristics and hence the switching times of the transistors were not need to be given much attention anymore, contrary to today's conditions~\cite{NAN_289}.

Besides the electrical characteristics changes, the quantum effects in MOS devices observed so far are relevant only for very low-temperature operation. 
It is still unknown whether further quantum effects occur below 30 nm channel lengths.

\subsection{Bipolar Transistors~\cite{NAN_2005}}
\begin{figure}
  \centering
  \includegraphics[width=7cm]{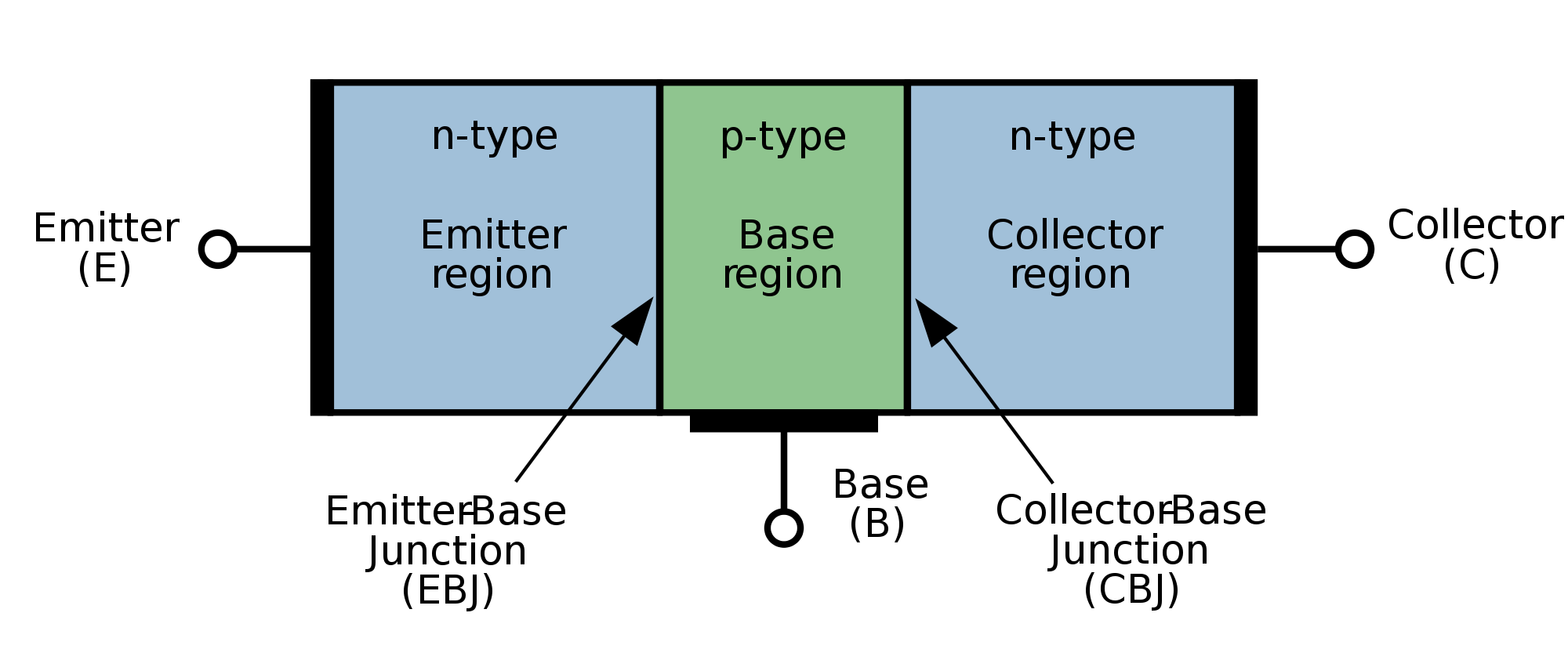}
  \caption{Junction section of NPN bipolar transistor~\cite{Bipolar}.}
  \label{Fig:Bipolar}
\end{figure}

The bipolar technology of transistors uses structures with nanometer dimensions only during the self-adjusting bipolar process. Due to the self-adjustment or self-assembly of the dopings relative to each other, the transit frequencies of bipolar integrated circuit technology could be enabled in the range above 40 GHz for pure silicon transistors and up to about 120 GHz for Silicon-Germanium(SiGe) switching elements. From the production aspects, extremely thin epitaxial films of different doping levels are used as the collector(100 nm) and base layers($<$ 50 nm) instead of implantation or diffusion. And, only the emitter is diffused from polysilicon layer into the crystal. Fig.~\ref{Fig:Bipolar} shows the junction section of NPN bipolar transistor.

The self-adjusting bipolar process is characterized by high critical frequencies($>$40 GHz) of the circuit elements in connection with a relatively high packing density. The typical area of the emitter amounts to about 0.15 $\cdot$ 1.5 $ \mu m^{2}$ in size. Also, the critical frequencies for further increases could be possible with a base layer from a heteroepitaxially grown crystalline silicon-germanium epitaxial layer which is deposited on a silicon substrate with the molecular beam epitaxy or via MOCVD (Metalorganic Chemical Vapor Deposition) procedure~\cite{MOCVD}.

Since many typical applications of the bipolar transistors in the high critical frequency regime are taken over today by MOS transistors, the fields of application of these elements in the future are exclusively used within the very high-frequency regime. And the heterojunction bipolar transistors from SiGe are particularly suitable for this purpose. The nanostructuring of bipolar transistors will also lead to a further increase in the critical frequencies, but no substantial technological innovation is to be expected in this area yet~\cite{NAN_2005}.

\subsection{Single Electron Transistors}
\begin{figure}
  \centering
  \includegraphics[width=7cm]{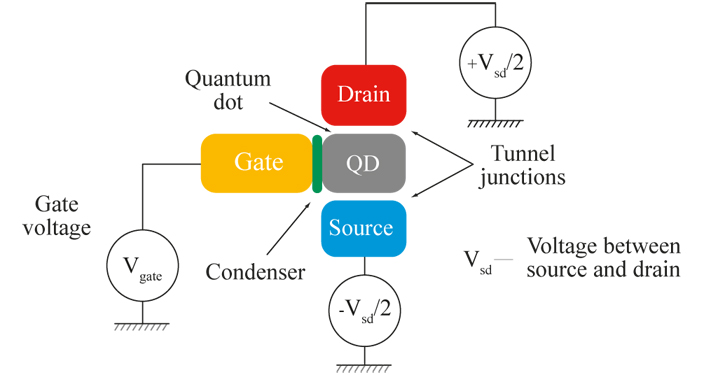}
  \caption{Schematic representation of the double barrier structure of a single electron transistor~\cite{NAN_359}.}
  \label{Fig:SET}
\end{figure}

The single electron transistor (SET) can be as an example of an electronic semiconductor device where a final limit of electronics has already been reached: the switching option of a current carried by just on the electron. The operating principle of SET can be more easily understood with the help of Fig.~\ref{Fig:SET}.

To further reduce the size of current devices beyond the limit of a hundred nanometers, the metal single electron transistor, using the Coulomb blocked effect, has been recently developed. The scaling of such nanosized devices down to atomic scaling can be expected to replace the customary semiconductor logic or analog devices. Due to the narrowing of the distance down to several nanometers between isolated electrodes such as the drain and source, the tunneling current readily surmounts conventional conduction currents. Thus, the tunneling of a single electron to the nanosized gate can build a high potential drop on account of the extremely smallness of the capacitance of the gate. Several constraints that limit the size of SET arise from the physical principles and device structure.

Many excellent reviews on the SET transistor~\cite{RAN_346}~\cite{RAN_347}~\cite{RAN_349} have been reported. In electron-beam lithography, the well-known proximity effect refers to variation in the width of patterned lines with the density of other shapes nearby, which this type variation, of course, makes increasing the resolution more difficult. Therefore, the electron proximity effect has been one of the major obstacles to achieving fine resolution in electron beam lithography. The distribution of intensity of exposure has a Gaussian intensity profile, because electrons are both forward scattered and back scattered. It can be partly compensated for the proximity effects by adjusting the dosage of electron beams according to the density of the patterns, or to anticipate the changes in dimensions of the features and then make compensating adjustments in advance.

In SETs, the nano-constriction between the source (or drain) electrode and the quantum dot of the Si-SET was formed by overlapping the distribution of the electron dosages of two separately written nano-dots performing on a silicon oxide insulating (SOI) wafer~\cite{RAN_346}. Also, bi-directional pump current, as well as single electrons transport, can be observed in the silicon dual-gate bi-directional electron pumps. The quantized current has been observed in a silicone dual-gate bi-directional electron pump. The polarity of the pump current can be altered either by the phase difference of the \textit{AC} modulations added to the gate voltages~\cite{RAN_349}, or by the \textit{DC} voltages applied to the two gate electrodes.

\subsection{Carbon Nanotube Transistors}
As demonstrated in Section~\ref{Sec:NF}, carbon nanotubes are made out of a structured network with the basic unit being six carbon atoms in the ring configuration and arranged in the form of cylinders. The electronic structure of the carbon nanotubes critically depends on its geometry of the interconnection between the carbon rings, resulting either in metallic or in semiconducting behavior~\cite{NAN_383}.

The particular interest in this new materials is due to reports of very low specific resistivities for metallic carbon nanotubes~\cite{NAN_388} and on high hole mobilities for semiconducting nanotubes~\cite{NAN_383}~\cite{NAN_387}. Those interesting electronic properties can be physically explained by the low density of the surface state. The material forms a two-dimensional network of carbon atoms without the presence of dangling bonds. When assembling in cylindrical form, the problem of the usually enhanced recombination at the edges of the semiconductor can be avoided~\cite{NAN_383}.

The small device dimensions in semiconducting carbon nanotubes together with the high values of the charge carrier mobilities make CNT-based devices very interesting for microelectronic applications. So far field-effect type transistors have mostly been implemented~\cite{NAN_389}~\cite{NAN_390}~\cite{NAN_391} because carbon nanotubes exhibit very high hole mobilities in particular. It should, however, also be mentioned that first experiments to realize a bipolar \textit{p-n-p} transistor were successful using CNTs transistors~\cite{NAN_393}.

One of the main problems regarding the fabrication of integrated circuits using CNT transistors is the limited reproducibility of the CNT growth process. An alternative approach to lateral integration is that manufacturing of arrays of CNTs based on vertical structures. Very homogeneous and reproducible growth of vertical CNT arrays by pyrolysis of acetylene on cobalt coated alumina substrates has already been reported~\cite{NAN_394}.

As a perspective for other applications of carbon nanotubes for electronic devices, it should be mentioned that heterojunctions between CNTs and silicon quantum wires have already been reported~\cite{NAN_397}. In this case, the silicon quantum wires are grown by CVD (Chemical Vapor Deposition)~\cite{MOCVD} deposition in a silane atmosphere selectively on top of the CNTs. They consist of a crystalline core covered by a thin amorphous silicon layer and a $SiO_{2}$ layer respectively. The electrical characterization of this heterostructure shows a behavior very similar to a Schottky diode and the current-voltage characteristics clearly exhibited rectifying behavior~\cite{RB_E}.

\subsection{Memristor}
\begin{figure}
  \centering
  \includegraphics[width=7cm]{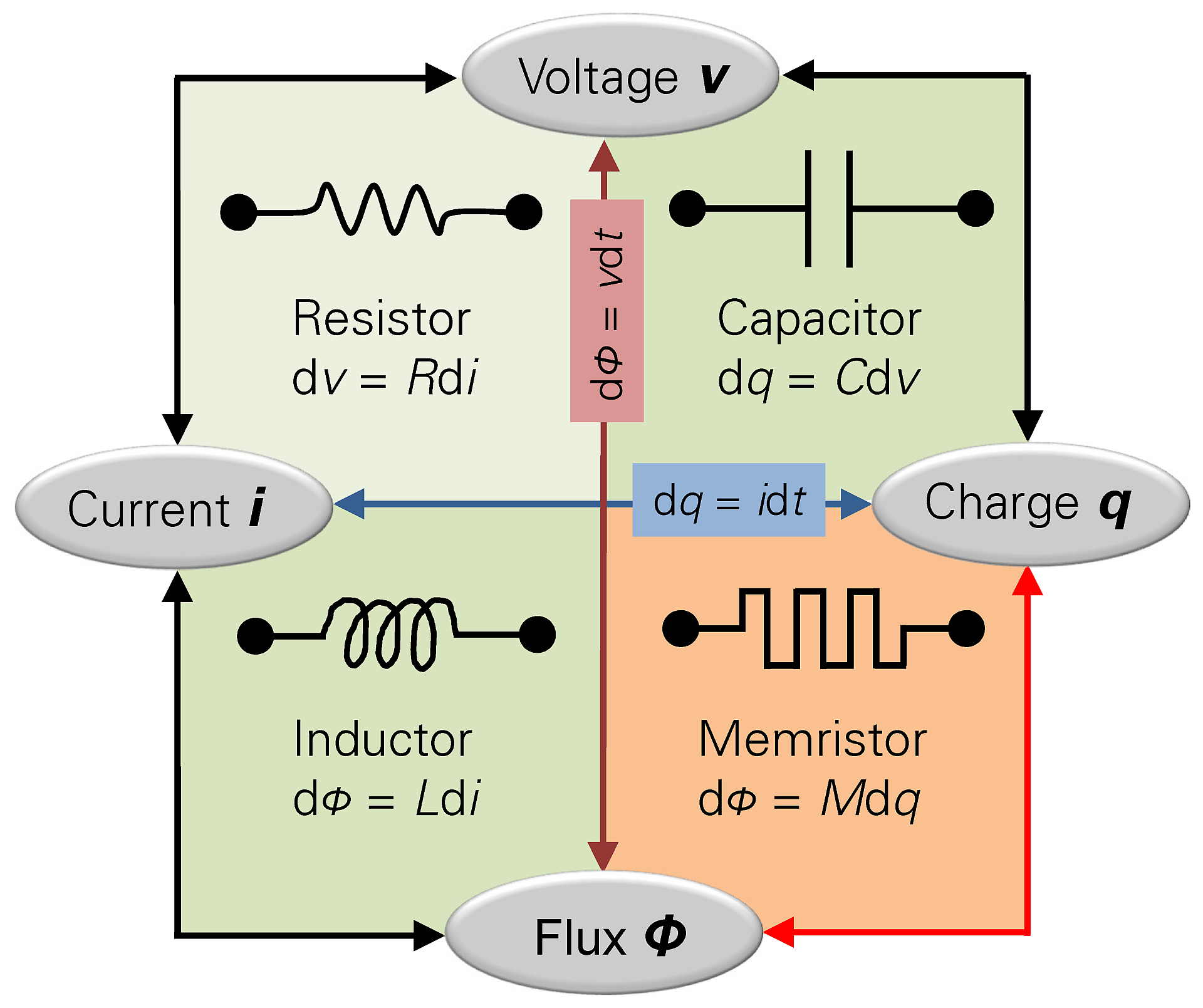}
  \caption{Illustration of the six combinations of the relationships between voltage $v$, charge $q$, flux $\varphi$, and current $i$. The memristor connects the charge and flux.~\cite{MER_PhD}} 
  \label{Fig:MER}
\end{figure}

Recent discovery of the memristor has sparked a new wave of enthusiasm and optimism in revolutionizing circuit design, marking a new era for the advancement of neuromorphic and analog applications. 
Leon Chua conceived the need for an additional fundamental circuit component in addition to the resistor, capacitor, and inductor~\cite{MER_P07}. Chua reasoned the existence of a missing circuit element from symmetry reasons, by looking at the six possible combinations of the relationships of four fundamental circuit variables - the voltage $V$, current $I$, flux $\varphi$, and electric charge $q$. While the charge is the integral upon the time of the current and the flux is integral upon the time of the voltage, the other possible relationships are connected by two-terminal circuit components. Resistors connect voltage to current by Ohm's law ($V = IR$), capacitors connect charge to voltage ($q = CV$), and inductors connect current to flux ($\varphi = LI$). The sixth possible relationship is the connection between charge and flux and is not covered by any conventional circuit element. Chua reasoned, for the sake of completeness, the existence of a fourth fundamental circuit element that connects the charge and flux and named the device the $memristor$, as a short for `$memory$ $resistor$'. The six combinations of the relationships are illustrated in Fig.~\ref{Fig:MER}.

The beauty of the memristor lies in its ability to remember its history state via the modulation of the internal state variables of the device. This memory capability is precisely what excites the electronics community and the underlying reason for the memristors revolutionary effects in circuit design. And, as CMOS technologies are approaching the nanoscale floor, Moore’s Law will eventually cease to exist, with devices attaining comparable dimensions to their constituting atoms. Thus, the focus has to be shifted to finding new devices which are increasingly infinitesimal and equally if not more capable than transistors. 

The memristor is a type of non-volatile memory. Digital applications usually require devices that combine long retention times with fast write speeds. The memristor can in practice achieve a long state lifetime ($\geq 10^7$) and ultra-fast switching speed, since relatively small biases can increase the switching speed up to six orders of magnitude due to the highly non-linear rate of switching. Additionally, the memristors' ability to maintain a state, without requiring external biasing, can significantly reduce the overall power consumption, while its deep nanoscale physical dimensions make the memristor (minimum reported: 5 x 5nm) an ideal candidate for implementing ultra-high-density memories, thus providing a much-needed extension to Moore’s law. As a consequence, memristors are often promoted as an emerging bi-stable switch for resistive random-access memory(RRAM).

Clearly, the characteristics of the scalability, the low power consumption and the dynamic response for the memristor are attributes that make this device attractive for a number of applications, from non-volatile memory~\cite{MER_R04} to programmable logic~\cite{MER_R05}. Particular emphasis is however given to the non-linear nature of memristor that resembles the behavior of chemical synapses~\cite{MER_R06}~\cite{MER_R07}, thus, marking a new era for neuromorphic engineering.
Memristors can be used in many applications, such as memory, logic, analog circuits, and neuromorphic systems. Also, memristors offer several outstanding advantages as compared to standard memory technologies: good scalability, non-volatility, effectively no leakage current, and compatibility with CMOS technology, both electrically and in terms of manufacturing~\cite{MER_Pp50}.

Although the device can be implemented as a bi-stable switch, its operation is not limited by discrete states, since a continuous memristive spectrum is attainable, signifying the potential employment of memristors as purely analog memory elements. Meanwhile, the memristance of a device can be varied in a very controlled manner by appropriate biasing operations. This property can be particularly useful in non-volatile memory applications where arbitrary signals can be utilized to program the conductance of device at multiple levels, thus increasing the memory capacity without increasing the number of elements~\cite{MEMR}. 

\subsection{PCM}
Over the past four decades, silicon technology has enabled data storage through charge retention on metal-oxide-silicon (MOS) capacitive structures. However, as silicon devices are continuously scaled toward (sub-) 10 nm dimensions, minute capacitors become very leaky by simple quantum mechanical considerations, thus the memory storage density appears to plateau.
Phase Change Memory(PCM) is an emerging technology which combines the unique properties of phase change materials with the potential for novel memory devices, which can further help lead to the new computer architectures. Phase change materials store information in their amorphous and crystalline phases respectively, which can be reversibly switched by the applying an external voltage~\cite{PCM}.

\begin{figure}
  \centering
  \includegraphics[width=9cm]{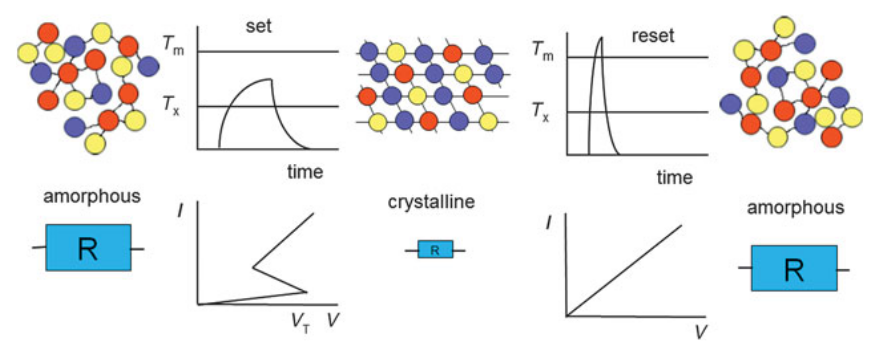}
  \caption{Principle of phase change memory. Starting from the amorphous phase with large resistance R, a current pulse is applied. At the threshold voltage $V_{T}$, the resistance drops suddenly, and a large current ($I$) flows that heats the material above the crystallization temperature $T_{x}$ for a sufficiently long time to crystallize (set operation). In the crystalline state, the resistance is low. A larger, short current pulse is applied to heat the material above the melting temperature $T_{m}$. The material is melt-quenched and returns to the amorphous, high resistance state (reset operation). In the schematic, different colors represent different atoms (such as Ge, Sb, and Te in the commonly used GeSbTe compounds) in the phase change materials.~\cite{PCM}.}
  \label{Fig:PCM}
\end{figure}

Phase change materials exist in an amorphous and one or sometimes several crystalline phases, and they can be rapidly and repeatedly switched between these phases by external activation, such as voltage. The switching can be typically induced by heating through optical pulses or electrical (Joule) heating. The optical and electronic properties can vary significantly between the amorphous and crystalline phases, and this combination of optical and electrical contrast and repeated switching allows data storage.
Today, many technologically useful phase change materials are chalcogenides, which owe their success in this regard to a unique combination of properties, which may include strong optical and electrical contrast, fast crystallization, and high crystallization temperature (typically several hundred degrees Celsius). Fig.~\ref{Fig:PCM} shows the general principles of phase change memory.

PCM mainly based on the repeated switching activities of phase change material between the amorphous and the crystalline states associated with a large change in resistance. Data information is stored in the phase of the material and is read by measuring the resistance of the PCM cell, meanwhile, the cell is programmed and read using electrical pulses~\cite{PCM}.

Phase change materials are at the heart of PCM technology, and their corresponding properties to a large extent determine its functionality and success. However, optimization of phase change materials is not only application specific but also technology node specific. For example, the threshold voltage is on the order of 1 V in current typical PCM cells, but if devices are scaled to much smaller dimensions, the threshold voltage scales with the size of the amorphous region, and for very small cells, it could become comparable to the read voltage such that every read operation could alter the cell state~\cite{PCM}.

Using PCM to replace DRAM is a formidable challenge, it is because very fast switching times in the nanoseconds range and extremely high cycle numbers of $\sim 10^{16}$ present a combination of requirements that have not been achieved by phase change materials nowadays. Also, DRAM replacement with PCM is a special case since DRAM is a volatile memory, whereas PCM is a non-volatile memory. They are not comparable. If PCM were to achieve DRAM-like performance, it would open up new possibilities to realize completely new computer architectures. Very fast switching times have been achieved for several phase change materials, including $Ge_2Sb_2Te_5$~\cite{PCM_27}~\cite{PCM_28} and GeTe~\cite{PCM_26} in actual PCM devices. However, the high cycle number remains an enormous challenge, and it appears that scaling to smaller dimensions of the phase change material is beneficial for cycling in some extent.

PCM cells cannot only be programmed in the on- or off-state, it can also be possible to reach intermediate resistance states. Up to 16 levels have been demonstrated using a write-and-verify scheme~\cite{PCM_60}. A continuous transition can be utilized between resistance levels in PCM devices in an analog manner, this effect can be used to program them to mimic the behavior of a synapse, for example. Such an attempt could lead to the design of a neuromorphic computer with electronic hardware that resembles the functions of brain elements, such as the neurons and synapses. The phenomenon of spike-timing-dependent plasticity (a biological process where the strength of connections between neurons are adjusted during learning) could be demonstrated in PCM devices using specific programming schemes~\cite{PCM_67}~\cite{PCM_70}. Image recognition using a neural network of PCM devices has also been demonstrated~\cite{PCM_71}~\cite{PCM_72}~\cite{PCM_73}. These could potentially lead to a compact and low power neuromorphic computing system that is capable of processing information through learning, adaptation, and probabilistic association like the brain~\cite{PCM}.

\subsection{Nanosensors}
One of the early applications of nanotechnology is in the field of nanosensors~\cite{EWN_100}~\cite{EWN_68}~\cite{EWN_31}~\cite{EWN_51}. A nanosensor is not necessarily a device merely reduced in size to a few nanometers, but a device that makes use of the unique properties of nanoparticles and nanomaterials to detect and measure new types of events appeared in the nanoscale~\cite{EWN_00}. However, there are no general rules about nanosensors with regards to their unique properties. Most reviews on nanosensors are focused on the particular type of sensors, such as biological nanosensors, optical nanosensors, and magnetic nanosensors, with many technical details involved. Here we present an overview of all nanosensors, showing similarities and fundamental differences among the various categories~\cite{ACR_00}. 

In most cases, nanosensors work together and need to communicate. Communication among nanosensors will expand the capabilities and applications of individual nano-devices both in terms of complexity and the range of operation. Each sensor has its sensing range technically. The detection range of existing nanosensors requires them to be inside the phenomenon that is being measured, and the area covered by a single nanosensor must be limited to its close sensing environment. However, a network of nanosensors can cover much larger areas and, with communication mechanism among them, perform additional in-network processing abilities. In addition, several existing nanoscale sensing technologies require the use of external excitation and measurement equipment to operate. Wireless communication between nanosensors and micro- and macro- devices will eliminate this need~\cite{EWN_00}. In this part, we mainly focus on nano-electromagnetic communication units to illustrate the general principles of nanosensors. 

\begin{figure}
  \centering
  \includegraphics[width=8cm]{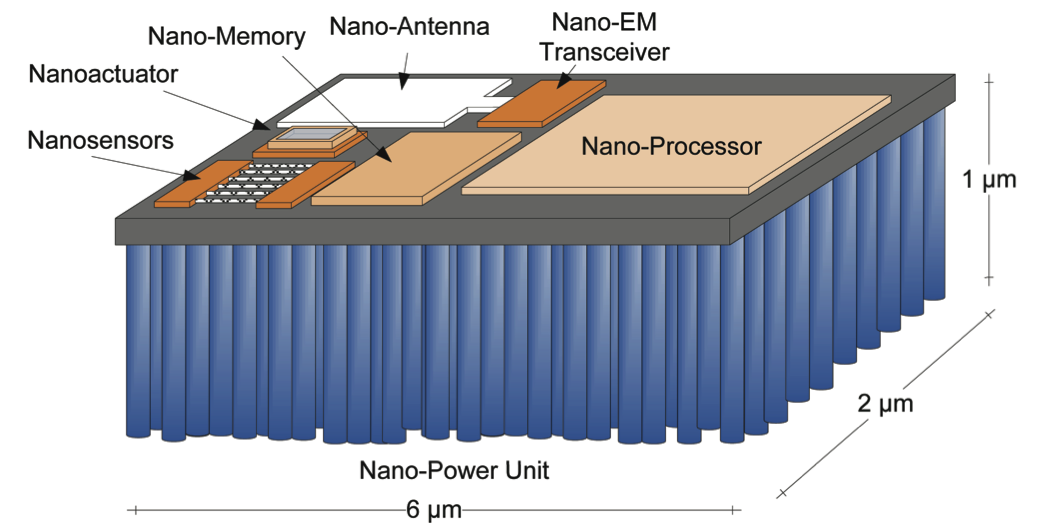}
  \caption{An integrated nanosensor device.~\cite{Web_NS}.}
  \label{Fig:NSensord}
\end{figure}

The nanosensors are as an integrated device around 10$\sim$100 $\mu m$ in size and able to do simple tasks besides sensing tasks, such as simple computation or even local actuation. The internal abstract architecture of a nanosensor device is shown in Fig.~\ref{Fig:NSensord}. In the abstract architecture, several important parts integrate a workable nanosensor device, such as sensing unit, actuation unit, power unit, a processing unit, storage unit, as well as a communication unit.

1. Sensing Unit

\begin{figure}
  \centering
  \includegraphics[width=8cm]{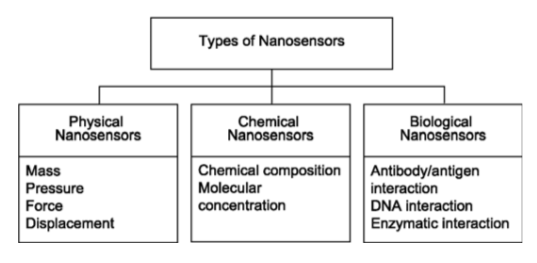}
  \caption{Types of nanosensors.~\cite{EWN_00}.}
  \label{Fig:NSensorty}
\end{figure}

Novel nanomaterials such as graphene and its derivatives, namely, Graphene Nanoribbons (GNRs) and Carbon Nanotubes (CNTs), provide outstanding sensing capabilities and are the basis for many types of sensors. Based on the nature of the different measured magnitudes, nanosensors can be classified as shown in Fig.~\ref{Fig:NSensorty}. Fig.~\ref{Fig:NSensorty} shows the types of state-of-the-art nanosensors, physical nanosensor, chemical nanosensor and biological nanosensor respectively, and their corresponding measure magnitudes.

2. Actuation Unit

An actuation unit will allow nanosensors to interact with their own close environment and can stimulate the simulation. Several nanoactuators have already been designed and implemented so far with outstanding actuation ability~\cite{EWN_51}. They can be classified as two types: Physical nanoactuators~\cite{EWN_51}~\cite{EWN_46}, and Chemical and biological nanoactuators~\cite{EWN_21}~\cite{EWN_38}.

However, the area of nanoactuators is at a very early stage compared with nanosensors technology. The main research challenge, besides the design and fabrication of the actuation unit, is how to precisely control and drive the nanoactuator and get the correct responses. The majority of potential applications of state-of-the-art nanosensors are used in the biomedical field; therefore, accuracy is one of the fundamental requirements for nanoactuators~\cite{EWN_00}.

3. Power Unit

As an integrated device, the power unit is used to provide energy supply for the whole nanosensor.  A major effort has been undertaken to reduce existing power sources to the microscale and the nanoscale. Nanomaterials can be used to manufacture nanobatteries with the outstanding advantages, such as high power density, reasonable lifetime and contained charge/discharge rates. However, having to periodically recharge them limits the usefulness of nanobatteries in realistic nanosensors’ applications and new technology is needed to overcome the power issues. 

In order to overcome the limitations of nanobatteries, the concept of self-powered nanodevices has been recently introduced in~\cite{EWN_90}~\cite{EWN_97}. The working principle of these self-powered devices is based on the conversion of the following types of energy into electrical energy which could power nanosensors:

$\bullet Mechanical \  energy$: produced for example by the human body movements, or muscle stretching.

$\bullet Vibrational \  energy:$ generated by acoustic waves or structural vibrations of buildings, among others.

$\bullet Hydraulic \  energy$: produced by body fluids, or the blood flow.

4. Processing Unit

Nanoscale processors are being enabled by the development of tinier FET transistors in different forms. Nanomaterials, such as CNTs and GNRs, can be used to build transistors in the range of nanometer scale~\cite{EWN_67}. The small size of nanosensor devices will limit the number of transistors in nanoscale processors, limiting the complexity of the operations that these will be able to do, but not the speed at which nano-processors will be able to operate~\cite{EWN_83}.

Independent of the specific approach followed to design these nano-transistors, one of the main challenges is in integrating them in future processor architectures. Although the experimental testing of individual transistors has been successfully conducted in literature, simple processing architectures based on these nano-transistors are still being investigated and, so far, the future processor architectures based on CNTs and graphene still need to be defined clearly before getting the actual prototypes.

5. Storage Unit

Ideally, nano-memories utilizing a single atom to store a single bit are being enabled by nanomaterials and new manufacturing processes~\cite{EWN_07}~\cite{EWN_59}. Several research challenges for nano-memories are summarized in twofold. First, for the time being, existing nanoscale memories require much more complex and expensive machinery to be written. Being able to read and write these memories in the nanoscale will be necessary for programmable nanosensor devices. Second, similarly to nano-processors, one of the main challenges is to mass manufacture compact nano-memories beyond simplified laboratory prototypes.

6. Communication Unit

Electromagnetic communication among nanosensors will be enabled and enhanced by the development of nano-antennas and the corresponding electromagnetic transceiver. 

$\bullet$ \textit{Nano-antennas}

When the antenna of a classical sensor device is reduced down to a few hundreds of nanometers, it would definitely require the use of extremely high operating frequencies, compromising the feasibility of electromagnetic wireless communication among nanosensor devices. However, the use of graphene as the material to fabricate nano-antennas can overcome this limitation. Indeed, the wave propagation velocity in CNTs and GNRs can be up to one hundred times below the speed of light in vacuum mainly depending on the structure geometry, temperature and Fermi energy~\cite{EWN_15}. Thus, the resonant frequency of nano-antennas based on graphene can be up to two orders of magnitude below that of nano-antennas built with non-carbon materials, so that the wave propagation velocity can be significantly improved.

$\bullet$ \textit{EM Nano-transceivers}

The EM (Embedded) transceiver of nanosensor device, embedded the necessary circuitry, can perform the baseband processing, frequency conversion, filtering, and power amplification, of the signals that have to be transmitted or that have been received from the free-space through the nano-antenna. Taking into account that the envisioned nano-antennas will resonate at frequencies in the terahertz band, RF FET transistors able to operate at these very high frequencies are necessary~\cite{EWN_52}~\cite{EWN_54}. 

The applications of wireless nanosensor networks can be classified into four main groups: biomedical, environmental, industrial, and military applications respectively. Wireless nanosensor networks will have a great impact in almost every field of our society, ranging from healthcare to homeland security and environmental protection. However, enabling the communication among nanosensors is still an unsolved challenge in nanotechnology areas.

\section{Discussions}
\label{Sec:Disc}
Nanotechnologies have added new capabilities to the state-of-the-art technologies and are projected to be commercially available in the near future. In the above sections, we have investigated these new properties and the implications of the new capabilities in the nano era. This combination of materials, physicals, semiconductors that are commonly used in nanodevices, are a disruptive technology, changing the way lives are organized today.
In this part, we will discuss several problems related nanotechnology, and follow with a potential prediction based on current developments.

\subsection{Problems with Nanotechnology}
When a new technology appeared, it usually sparks conflicts between those wishing to exploit it as rapidly as possible and those wishing to wait--forever, if necessary--to have it proved absolutely safe. Nanotechnology is relatively new compared with other technologies; although parts of it are quite familiar now, parts are unfamiliar, and it is not a surprise that the public is wary of its potential for harm, as well as excited by its potential for good~\cite{White_2005}.

\subsubsection{Uncontrollability}
Nanotechnology has changed and will continue to change our vision, expectations, and abilities to control the materials world. These developments in nanoscale will definitely affect the physical and chemical properties of materials. Recent major achievements include the ability to observe structure at its atomic level and measure the strength and hardness of microscopic and nanoscopic phases of composite materials.

However, one concern is that nanotechnology could go out of control if it is not correctly used. This concern is based on an idea put forward by several futurists (Drexler, Joy, and others)~\cite{White_22}~\cite{White_24}, and adopted gleefully by science fiction writers~\cite{White_23}: that is, the idea of small machines that can replicate themselves automatically (``assemblers”) and that escape from the laboratory and eat the earth. Any statement about the future is, of course, always personal opinion. But there is no way that such devices can exist on earth. The idea of small, self-replicating machines does not seem impossible now —- after all, bacteria exist -— but developing such machines de novo -— a task close to developing a new form of life -— has seemed to the public to be intractably difficult; it continues to seem so. Self-replicating nanomachines that resemble the larger machines with which we are familiar can be built. So, this type of concern can be dismissed, at least until and unless scientific inventions —- in self-replication, and in artificial life -— appear that will far exceed nanoscience in their importance~\cite{White_2005}.

A promising and special direction of research in the field of nanoelectronic devices is to create storage devices having super high density and terabit capacity using nanomaterials. The idea that such storage devices can be created appeared soon after STM (STMicroelectronics) capable of manipulating separate atoms has been developed. To meet the requirements of this type storage, the concept of using self-organizing ordered atomic molecular structures as a storage medium has been put forward~\cite{NMA}. Some progress for this type storage has achieved. Ordered structures of organosilicon compounds on graphite substrates have been tested, and the possibility of recording memory elements of size 0.5 nm have been demonstrated. However, the uncontrollability of the STM probe tip at the atomic level and the limited set of substrates allowing work under normal conditions have yielded no way of going beyond separate successful experiments. Moreover, work performed under conditions of super-high vacuum on atomically clean surfaces have shown that carrying separate atoms from a substrate to a probe or from a probe onto a substrate is far from being simple enough to be achieved using state-of-the-art technologies. These processes are of probabilistic character with characteristic times at the level of several to tens of milliseconds, so the necessity for normal storage operation recording validity at level $10^5 - 10^6$ is out of the question at the present stage.

\subsubsection{Health}
It found that most nanotechnologies pose no new risks to humans or the environment. However, much more work need to be done before research can say how dangerous these nanoparticles could be. For example, it is unclear that nanoparticles would do if they entered the human body. Micrometer-sized clumps of nanoparticles, for example, are relatively unreactive because their surface areas are smaller than that of the same number of individual nanoparticles, and they are too large to enter the blood stream when breathed in. But individual nanoparticles can pass from the lungs into the bloodstream, and are more reactive~\cite{NAL}.

Another issue is the unknown toxicity of materials to health. Materials can behave quite differently at the nanoscale to the way they do in bulk. This is both because the small size of the particles dramatically increases surface area and therefore reactivity, and also because quantum effects start to become significant. This potential difference is just what makes them interesting to scientists and engineers. However, it also means that their toxicity may be different from that of the same chemical in the form of larger particles~\cite{DON_6}~\cite{DON_7}. There are examples where nanoparticles can produce toxic effects even if the bulk substance is nonpoisonous. This arises partly because they have increased surface area and also because, should the nanoparticles enter the body through inhalation, ingestion, or absorption through the skin, they are able to move around and enter cells more easily than larger particles.

Little research has been carried out on the toxicity of manufactured nanoparticles and nanotubes, but we can learn from studies on the effects of exposure to mineral dust in some workplaces and to the nanoparticles in air pollution. Considerable evidence from industrial exposure to mineral dust demonstrates that the toxic hazard is related to the surface area of the inhaled particles and to their surface activity. Epidemiological studies of urban air pollution support the conclusion that finer particles cause more harm than coarser ones – diesel $PM_{10}$ pollution is implicated in heart and lung disease and asthma, particularly in susceptible people. Although we breathe in millions of pollutant particles with each breath, apparently without serious harm, increases of only 10 $\mu g/m^3$ are consistently associated with a 1\% increase in cardiac deaths.

It is very unlikely that new, manufactured nanoparticles could be introduced into humans in doses sufficient to cause the health effects that have been associated with air pollution. However, some may be inhaled in certain workplaces in significant amounts and steps should be taken to minimize exposure. Toxicological studies have investigated nanoparticles of low solubility and surface activity. Newer particles with characteristics that differ substantially from these should be treated with particular caution.

Long, thin fibers like asbestos (narrower than about 3 $\mu m$ and longer than about 15 $\mu m$) are a particular cause for concern~\cite{DON_8}. They have aerodynamic properties that allow them to reach the gas-exchanging part of the lung when inhaled, but are too long to be removed by macro-phages, the lung’s scavengers. Once lodged deep in the lungs they can inflame the tissue and may eventually lead to scarring and lung cancer. We have concerns about carbon nanotubes, which could conceivably cause similar problems to asbestos if inhaled in quantity as single fibers. Current manufacturing techniques tend to lead to nanotubes that are clumped into ‘bundles’. However, much current activity is directed toward developing techniques and coatings to enable the nanotubes to remain separate. A successful outcome to that research will lead to many more applications for nanotubes, but will also mean that they may readily become airborne and inhaled. As the nanotubes are designed to be insoluble, they may remain in the lung tissue and induce the free-radical release that produces inflammation. Until further toxicological studies have been undertaken, human exposure to airborne nanotubes in laboratories and workplaces should be minimized~\cite{DON}.

Here public concern has a legitimate basis. We do not, in fact, understand the interaction of small particles with cells and tissues, but there are diseases associated with a few of them: silicosis, asbestosis, “black lung”~\cite{White_73}~\cite{White_74}. Most nanomaterials are probably safe: there is no reason to expect fundamentally new kinds of toxicity from them, and in any event, they are common in the environment. Moreover, in commerce, most would be made and used in conditions in which the nanomaterial was relatively shielded from exposure to society (an example would be buckytubes compounded into plastics). Still, we do not know how nanoparticles enter the body, how they are taken up by the cell, how they are distributed in the circulation, or how they affect the health of the organism. If the chemical industry intends to make a serious entry into nanostructured materials, it would be well advised to sponsor arms-length, careful, and entirely dispassionate studies on the effects of existing and new nanoparticles and nanomaterials on the behavior of cells and on the health of animals. This particular aspect of public health will, in any event, be examined in detail by regulatory agencies concerned with the effects of nanoparticulate from other sources (especially carbon nanoparticles in the exhaust from diesel engines) on health.

\subsubsection{Privacy and Ethical}
The most serious risk of nanotechnology comes, not from hypothetical revolutionary materials or systems, but from the uses of evolutionary nanotechnologies that are already developing rapidly. The continuing extension of electronics and telecommunications —- fast processors, ultradense memory, methods for searching databases, ubiquitous sensors, electronic commerce and banking, commercial and governmental record keeping —- into most aspects of life is increasingly making it possible to collect, store, and sort enormous quantities of data about people~\cite{White_75}. These data can be used to identify and characterize individuals, and the ease with which they can be collected and manipulated poses a direct threat to historical norms of individual privacy. “Universal surveillance” -— the observation of everyone and everything, in real time, everywhere; a concept suggested by those most concerned with terrorism -— is not a technology that we would wish to see cloak a free society, no matter how protectively intended~\cite{White_76}.

The risk of new information technologies emerges naturally and almost invisibly from an existing technology with which society is already comfortably familiar, and in which there is no fundamentally ``new" concept, and nothing uniquely associated with ``nano". There is, however, no question that information technology has already (and to a far greater extent that biotechnology) transformed the world. I believe that it will continue to do so, and that transformation is more pervasive and deep-seated than anything that will come from ``revolutionary" nanotechnology in the foreseeable future.

In the short term, the societal concerns that arise in the development of nanotechnologies are centered around `who controls the uses of nanotechnologies' and `who benefits from uses of nanotechnologies'. Similar concerns apply to any new technology.

In the longer term, the convergence of nanotechnology with other technologies is expected to lead to far-reaching developments, which may raise social and ethical issues. The convergence of nanotechnology with biotechnology, information, and cognitive sciences may lead to artificial retinas and so help the blind to see, but more radical forms of human enhancement have been postulated~\cite{DON_12}, which, if feasible, would raise profound ethical issues. A number of the social and ethical issues that might be generated by nanotechnologies should be investigated further, so we recommend the establishment of a multidisciplinary research program to do this. The ethical and social implications of advanced technologies should form part of the formal training of all research students and staff working in these areas.

Public attitudes can play a crucial role in realizing the potential of technological advances. 

\subsection{Prediction}
This part predicts the current development trends from the information and technology aspects considering the state-of-the-art nanotechnologies.

\subsubsection{Sustainable Nanomaterials}
Nanotechnology is among the most prominent emerging technologies and it heralded as a key technology for the 21st century. Potential innovations offer numerous benefits. There are great expectations among policymakers, scientists and industry representatives that nanotechnology may or will contribute to economic prosperity and sustainable development (for an up-to-date and comprehensive overview see Ref.~\cite{P_NFE}. On the other hand, nanotechnology has been the subject of an extensive public debate in Europe and the United States. 
Obviously, nanotechnology is a case for technology assessment$~\cite{P_AET}$.

The segment of `nanotechnology' that is closest to a widespread application is the field of 'nanomaterials'. Nanomaterials are an essential part of the overall field of nanotechnology. They can be considered as the most important bridge between basic research and marketable products and processes. As so-called, `enabling technologies', they are technological prerequisites for numerous innovations in many technological fields from comparatively simple technologies for every day use (like cosmetics or pigments in paints), energy technologies or information and communication up to biotechnologies -- without their interdependence being always obvious at first glance. Some nanomaterials -- based products and processes are already in the marketplace, much more will very likely be seen in the near or mid-term future$~\cite{P_AET}$. 

Nanomaterials show great economic potential, e.g. by substituting other materials or by making available new functionalities and thus enabling new products and creating new markets. It is also expected that nanomaterials may contribute to the reduction of the ecological footprint of classical production processes by reducing energy and material consumption.

There continues to be a lack of complete understanding regarding the environmental, health, and safety (EHS) effects of exposure to engineered nanoscale materials, governments, industry, and other stakeholders are considering how best to address EHS issues while continuing to foster the sustainable commercialization of nanoscale materials. It is generally believed that sufficient information exists about the toxicity of some nanoscale materials to suggest a need for caution. The small size of certain nanoparticles facilitates their uptake into cells and their movement through the body more readily than is the case with their conventionally sized counterparts~\cite{P_NAE}. Other factors contribute to a sense of uncertainty as to the biological and environmental implications of exposure to nanoscale materials. Size, shape, surface chemistry, and coating, for example, can all influence how these materials behave biologically and in the environment. The fact that nanoscale materials can have unusual properties, properties that do not conform to ``conventional” physics and chemistry, may increase their commercial value and their potential risks~\cite{P_SNE}.

Nano products are diverse and growing exponentially. According to the National Nanotechnology Initiative (NNI), nanoscale materials are used in electronics, pharmaceuticals, chemicals, energy, and biomedical, among other industries. These products include paints, sporting goods, cosmetics, stain-resistant clothing, electronics, and surface coatings, among other applications~\cite{P_ANC}.

Sustainable technologies are, in our view, characterized by high benefits, low risks for the short- and long-term and social acceptance. It is important to recognize that technologies are not invented in a vacuum, but emerge from the interplay within a wide constellation of societal activities and actors~\cite{P_NAS}. Technologies are therefore, indeed, a product of societal systems. Sustainability has become an umbrella term for many different things. While in most approaches environmental concerns are highlighted, as well economical and social aspects are stressed. Various definitions of sustainability circulate, committees struggle about the adequate application of the term and consultants offer ingenious indicators~\cite{P_DON}.

Enter green nanotechnology, a conceptual approach to managing EHS risks potentially posed by nanoscale materials to ensure their responsible and sustainable development. There are two key aspects of green nanotechnology. The first involves nanoproducts that provide solutions to environmental challenges. These include environmental technologies to remediate hazardous waste sites and desalinate water, nanotechnology applications for improving food nutritional value, nanoproducts that facilitate sensing and monitoring technologies to detect hazardous pollutants, and other applications. The second involves producing nanomaterials and nano enabled products in ways that minimize human and environmental harm. New nanomaterials can be made using well-established principles of green chemistry, thus avoiding dependence on processes that might result in pollutants.

Green engineering principles are applicable as nanomaterials increasingly are incorporated into larger, more conventionally scaled products. Green engineering embraces the concept that decisions to protect human health and the environment can have the greatest impact and cost effectiveness when applied early to the design and development phase of a process or product. The most relevant time frame in the green engineering lifecycle of a nanomaterial is the design stage. Green engineering considers the full lifecycle of a product, from the extraction of the materials through manufacturing, product use, and end of life. Green nanotechnologies which focus on the full lifecycle can better prepare users for recycling, reuse, or remanufacture of nanomaterials and nano-enabled products, thus minimizing generating new hazards through unintended consequences.

Nanomaterials can be designed to be sustainable. Nanomaterials can, for example, be coated so that they do not dissolve in water or enter biological cells. Some nanomaterials can be made from renewable ingredients or repurpose nontoxic biological waste products. Other nanomaterials can be considered to ensure no part of the product can be the source of harm to human or environmental health after gainful use and reclamation opportunities are exhausted.

A subset of greener production includes using nanomaterials to ``green up" current processes. Catalysts are an important nanomaterial for this use. As a spherical particle gets smaller and smaller, it has more surface area proportional to its total volume. Catalyst reactions take place on the surface, so the more surface area and less volume the better. Nanomaterials used as catalysts have high surface areas making them more efficient and less wasteful, with potentially less polluting chemical reactions.

Nanoscale membranes are another illustration of green nano applications. In many chemical reactions, useful products must be separated from waste. These separations can be energy intensive, wasteful, or themselves polluting. Nanoscale membranes can minimize separation steps and energy use.

These examples are merely illustrative of a broad range of green nanoproducts and processes. While there is reason to be hopeful, there is also reason to be cautious when creating and managing these new, unique materials and manufacturing processes.

\subsubsection{Nano-Circuits}
Nanocircuits are electrical circuits operating on the nanometer scale, which is well into the quantum realm, and quantum mechanical effects become very important.  A variety of proposals have been made to implement nano-circuits, including Nanowires, single-election transistors, quantum dot cellular automate, and nanoscale crossbar latches. 

Taking the three-dimensional integrated circuit (3D IC) as an example, 3D IC is an integrated circuit manufactured by stacking silicon wafers or dies and interconnecting them vertically, e.g., using through-silicon vias (TSVx), so that they behave as a single device to achieve performance improvements at reduced power and smaller footprint than conventional two dimensional processes. Stacking is important in 3D IC.  There exist many key stacking approaches being implemented and explored, including die-to-die, die-to-wafer and wafer-to-wafer. However, these technologies are not mature, and they carry new challenges, such as cost, yield, design complexity, TSV-introduced overhead, lack of standards, etc. Thus, these challenges urgently require new technology to deal with. Nanotechnology provides a good solution for these challenges. 


\section{Conclusion}
\label{Sec:Conc}
Nanotechnologies offer great opportunities and continue to attract a lot of attention because of their potential impacts on an incredibly wide range of industries and markets. Consequently, this technology is evolving rapidly and will develop faster over the coming years. The potential new features of nanotechnology will be to promote developing the new nanodevices. Meanwhile, it is also essential to address uncertainties and the potential problems which nanotechnologies may take in an economic and safe manner.

\end{document}